\documentclass[12pt,titlepage,letterpaper]{utarticle}

\usepackage{amsmath}
\usepackage{mathrsfs}
\usepackage{amsfonts}
\usepackage{amssymb}
\usepackage{amsthm}
\usepackage{mathtools}
\usepackage{mathbbol}
\usepackage{graphicx}
\usepackage{color}
\usepackage{xparse}
\usepackage{tikz}
\usepackage{tocloft}
\usepackage[titletoc,title]{appendix}
\usepackage[nosort]{cite}
\usepackage{hyperref}
\usepackage{longtable}

\usepackage{caption}

\definecolor{lightmauve}{RGB}{255,187,255}
\definecolor{lightblue}{RGB}{238,238,255}
\definecolor{lightred}{RGB}{255,238,238}
\definecolor{midgreen}{RGB}{15,170,15}

\definecolor{aqua}{rgb}{0, 1.0, 1.0}
\definecolor{fuschia}{rgb}{1.0, 0, 1.0}
\definecolor{gray}{rgb}{0.502, 0.502, 0.502}
\definecolor{lime}{rgb}{0, 1.0, 0}
\definecolor{maroon}{rgb}{0.502, 0, 0}
\definecolor{navy}{rgb}{0, 0, 0.502}
\definecolor{olive}{rgb}{0.502, 0.502, 0}
\definecolor{purple}{rgb}{0.502, 0, 0.502}
\definecolor{silver}{rgb}{0.753, 0.753, 0.753}
\definecolor{teal}{rgb}{0, 0.502, 0.502}

\makeatletter
\newdimen\itex@wd%
\newdimen\itex@dp%
\newdimen\itex@thd%
\def\itexspace#1#2#3{\itex@wd=#3em%
\itex@wd=0.1\itex@wd%
\itex@dp=#2ex%
\itex@dp=0.1\itex@dp%
\itex@thd=#1ex%
\itex@thd=0.1\itex@thd%
\advance\itex@thd\the\itex@dp%
\makebox[\the\itex@wd]{\rule[-\the\itex@dp]{0cm}{\the\itex@thd}}}
\makeatother

\makeatletter
\newif\if@sup
\newtoks\@sups
\def\append@sup#1{\edef\act{\noexpand\@sups={\the\@sups #1}}\act}%
\def\reset@sup{\@supfalse\@sups={}}%
\def\mk@scripts#1#2{\if #2/ \if@sup ^{\the\@sups}\fi \else%
  \ifx #1_ \if@sup ^{\the\@sups}\reset@sup \fi {}_{#2}%
  \else \append@sup#2 \@suptrue \fi%
  \expandafter\mk@scripts\fi}
\def\tensor#1#2{\reset@sup#1\mk@scripts#2_/}
\def\multiscripts#1#2#3{\reset@sup{}\mk@scripts#1_/#2%
  \reset@sup\mk@scripts#3_/}
\makeatother

\makeatletter
\newbox\slashbox \setbox\slashbox=\hbox{$/$}
\def\itex@pslash#1{\setbox\@tempboxa=\hbox{$#1$}
  \@tempdima=0.5\wd\slashbox \advance\@tempdima 0.5\wd\@tempboxa
  \copy\slashbox \kern-\@tempdima \box\@tempboxa}
\def\slash{\protect\itex@pslash}
\makeatother

\def\clap#1{\hbox to 0pt{\hss#1\hss}}

\let\oldroot\root
\def\root#1#2{\oldroot #1 \of{#2}}
\renewcommand{\sqrt}[2][]{\oldroot #1 \of{#2}}

\DeclareSymbolFont{symbolsC}{U}{txsyc}{m}{n}
\SetSymbolFont{symbolsC}{bold}{U}{txsyc}{bx}{n}
\DeclareFontSubstitution{U}{txsyc}{m}{n}

\DeclareSymbolFont{stmry}{U}{stmry}{m}{n}
\SetSymbolFont{stmry}{bold}{U}{stmry}{b}{n}

\DeclareFontFamily{OMX}{MnSymbolE}{}
\DeclareSymbolFont{mnomx}{OMX}{MnSymbolE}{m}{n}
\SetSymbolFont{mnomx}{bold}{OMX}{MnSymbolE}{b}{n}
\DeclareFontShape{OMX}{MnSymbolE}{m}{n}{
    <-6>  MnSymbolE5
   <6-7>  MnSymbolE6
   <7-8>  MnSymbolE7
   <8-9>  MnSymbolE8
   <9-10> MnSymbolE9
  <10-12> MnSymbolE10
  <12->   MnSymbolE12}{}

\makeatletter
\def\re@DeclareMathSymbol#1#2#3#4{%
    \let#1=\undefined
    \DeclareMathSymbol{#1}{#2}{#3}{#4}}
\re@DeclareMathSymbol{\neArrow}{\mathrel}{symbolsC}{116}
\re@DeclareMathSymbol{\neArr}{\mathrel}{symbolsC}{116}
\re@DeclareMathSymbol{\seArrow}{\mathrel}{symbolsC}{117}
\re@DeclareMathSymbol{\seArr}{\mathrel}{symbolsC}{117}
\re@DeclareMathSymbol{\nwArrow}{\mathrel}{symbolsC}{118}
\re@DeclareMathSymbol{\nwArr}{\mathrel}{symbolsC}{118}
\re@DeclareMathSymbol{\swArrow}{\mathrel}{symbolsC}{119}
\re@DeclareMathSymbol{\swArr}{\mathrel}{symbolsC}{119}
\re@DeclareMathSymbol{\nequiv}{\mathrel}{symbolsC}{46}
\re@DeclareMathSymbol{\Perp}{\mathrel}{symbolsC}{121}
\re@DeclareMathSymbol{\Vbar}{\mathrel}{symbolsC}{121}
\re@DeclareMathSymbol{\sslash}{\mathrel}{stmry}{12}
\re@DeclareMathSymbol{\boxslash}{\mathrel}{stmry}{27}
\re@DeclareMathSymbol{\boxbslash}{\mathrel}{stmry}{28}
\re@DeclareMathSymbol{\boxast}{\mathrel}{stmry}{24}
\re@DeclareMathSymbol{\boxcircle}{\mathrel}{stmry}{29}
\re@DeclareMathSymbol{\boxbox}{\mathrel}{stmry}{30}
\re@DeclareMathSymbol{\obslash}{\mathrel}{stmry}{20}
\re@DeclareMathSymbol{\obar}{\mathrel}{stmry}{58}
\re@DeclareMathSymbol{\olessthan}{\mathrel}{stmry}{60}
\re@DeclareMathSymbol{\ogreaterthan}{\mathrel}{stmry}{61}
\re@DeclareMathSymbol{\bigsqcap}{\mathop}{stmry}{"64}
\re@DeclareMathSymbol{\biginterleave}{\mathop}{stmry}{"6}
\re@DeclareMathSymbol{\invamp}{\mathrel}{symbolsC}{77}
\re@DeclareMathSymbol{\parr}{\mathrel}{symbolsC}{77}
\makeatother

\makeatletter
\def\Decl@Mn@Delim#1#2#3#4{%
  \if\relax\noexpand#1%
    \let#1\undefined
  \fi
  \DeclareMathDelimiter{#1}{#2}{#3}{#4}{#3}{#4}}
\def\Decl@Mn@Open#1#2#3{\Decl@Mn@Delim{#1}{\mathopen}{#2}{#3}}
\def\Decl@Mn@Close#1#2#3{\Decl@Mn@Delim{#1}{\mathclose}{#2}{#3}}
\Decl@Mn@Open{\llangle}{mnomx}{'164}
\Decl@Mn@Close{\rrangle}{mnomx}{'171}
\Decl@Mn@Open{\lmoustache}{mnomx}{'245}
\Decl@Mn@Close{\rmoustache}{mnomx}{'244}
\Decl@Mn@Open{\llbracket}{stmry}{'112}
\Decl@Mn@Close{\rrbracket}{stmry}{'113}
\makeatother

\makeatletter
\DeclareRobustCommand\widecheck[1]{{\mathpalette\@widecheck{#1}}}
\def\@widecheck#1#2{%
    \setbox\z@\hbox{\m@th$#1#2$}%
    \setbox\tw@\hbox{\m@th$#1%
       \widehat{%
          \vrule\@width\z@\@height\ht\z@
          \vrule\@height\z@\@width\wd\z@}$}%
    \dp\tw@-\ht\z@
    \@tempdima\ht\z@ \advance\@tempdima2\ht\tw@ \divide\@tempdima\thr@@
    \setbox\tw@\hbox{%
       \raise\@tempdima\hbox{\scalebox{1}[-1]{\lower\@tempdima\box
\tw@}}}%
    {\ooalign{\box\tw@ \cr \box\z@}}}
\makeatother

\makeatletter
\NewDocumentCommand\mathraisebox{moom}{%
\IfNoValueTF{#2}{\def\@temp##1##2{\raisebox{#1}{$\m@th##1##2$}}}{%
\IfNoValueTF{#3}{\def\@temp##1##2{\raisebox{#1}[#2]{$\m@th##1##2$}}%
}{\def\@temp##1##2{\raisebox{#1}[#2][#3]{$\m@th##1##2$}}}}%
\mathpalette\@temp{#4}}
\makeatletter

\makeatletter
\def\udots{\mathinner{\mkern2mu\raise\p@\hbox{.}
\mkern2mu\raise4\p@\hbox{.}\mkern1mu
\raise7\p@\vbox{\kern7\p@\hbox{.}}\mkern1mu}}
\makeatother

\theoremstyle{plain}

\theoremstyle{definition}

\theoremstyle{remark}

\usetikzlibrary{positioning,positioning,arrows,decorations.markings}

\begin{document}

\preprint{
UTWI--18--2022\\
}

\title{Current Algebra Levels in the $E_8$ Theory}

\author{Jacques Distler, Grant Elliot
     \oneaddress{
      Weinberg Institute for Theoretical Physics\\
      Department of Physics,\\
      University of Texas at Austin,\\
      Austin, TX 78712, USA \\
      {~}\\
      \email{distler@golem.ph.utexas.edu}\\
      \email{gelliot123@utexas.edu}
      }
}
\date{December 6, 2022}

\Abstract{
The class-S theories of type $E_8$ were analyzed in \cite{Chacaltana:2018vhp}. The basic building blocks  consisted of 49,836  isolated SCFTs (three-punctured spheres). In 244 cases, there were undetermined levels for the flavour symmetry current algebra. Here, we rectify that omission.  Using S-duality and nilpotent Higgsing we compute the levels of 235 of the 244. There remain 9 three-punctured spheres with unknown levels. Along the way, we provide a detailed discussion of the Drinfeld-Sokolov reduction of the VOA, which captures various features of the Higgs branch RG flows arising from giving a VEV to a  nilpotent moment map.
}

\maketitle

\tocloftpagestyle{empty}
\tableofcontents
\vfill
\newpage
\setcounter{page}{1}

\section{Introduction}\label{introduction}

The 6D (2,0) theories, and hence the class-S theories that arise as their compactifications on punctured Riemann surfaces, have an ADE classification. The most formidable of these is, naturally, the $E_8$ theory. Its basic building blocks consist of 49,836 three-punctured spheres (``fixtures"), whose basic properties were computed in  \cite{Chacaltana:2018vhp}. Among these basic properties are the current-algebra levels for every simple factor in the flavour symmetry algebra of the SCFT. In the analysis of \cite{Chacaltana:2018vhp},  244 fixtures, out of 49836,  had current algebra levels that the authors of \cite{Chacaltana:2018vhp} were unable to determine. 

In this work, we rectify that deficiency, determining the previously unknown levels for 235 of the 244 fixtures. Our main tool is the Higgs branch RG flows induced by turning on a VEV for the highest root moment map for some simple factor in the flavour symmetry associated to a puncture.

These ``nilpotent Higgsings" have a realization in the 2D VOA  associated to the $\mathcal{N}=2$ SCFT as Drinfeld-Sokolov reduction \cite{Beem:2019snk}. We discuss that DS reduction in detail in \S\ref{DSreduction}. Since the DS reduction manifestly preserves the \emph{rest} of the flavour symmetry algebra (and its levels), it allows us to track those levels through the RG flow. We previously applied similar techniques to class-S theories of types $E_6$, $E_7$ in \cite{Distler:2022nsn}. In the $E_8$ theory, a few of our results  --- when one of the punctures is the simple puncture, $E_8(a_1)$ (entries 158--167 and 224--230 in Table \ref{fixedfixtures} below) --- have previously been determined by the methods of \cite{Baume:2021qho}. In those cases, our results agree with theirs.

\section{Determining Unknown Current-Algebra Levels}\label{DeterminingLevels}

Fixtures are class-S theories obtained by compactifying the (2,0) theory on a sphere with three punctures. Each puncture has an associated flavour symmetry and the fixture has a flavour symmetry which is (possibly an enhancement of) the product of the flavour symmetries associated to each puncture. The latter (the ``manifest" flavour symmetry) embeds as a subalgebra of the full flavour symmetry. The flavour symmetries of fixtures obtained from the $E_8$ (2,0) theories were determined in \cite{Chacaltana:2018vhp}.

To each simple flavour symmetry factor one may associate a positive integer $k$ that is the flavour central charge or ``level". The flavour central charge of a simple factor of the flavour symmetry is defined by
\begin{displaymath}
J^{a}_{\mu}(x)J^{b}_{\nu}(0) \sim \frac{3k}{4 \pi ^4} \delta^{ab}\frac{x^2 g_{\mu \nu}-2x_{\mu}x_{\nu}}{x^8}+\frac{2}{\pi^2}f^{abc}\frac{x_{\mu}x_{\nu}x \cdot J^{c}(0)}{x^6}+\dots
\end{displaymath}
where the normalization is such that $k=1$ for a free half-hypermultiplet in the defining representation of $Sp(n)$. The levels of the manifest flavour symmetries are readily determined from the decomposition of the adjoint representation \cite{Chacaltana:2012zy}, as described in \S2.4.1 of \cite{Chacaltana:2014jba}.

In most cases, when the flavour symmetry is enhanced, knowing the levels of the ``manifest" subalgebra suffices to determine the levels of the full flavour symmetry. There are, however, two notable exceptions.
\begin{itemize}
\item When a manifest factor of $G_k$ (we denote the level $k$ by a subscript) is enhanced to $G_{k_1}\times G_{k_2}$, where $G$ is embedded diagonally in $G\times G$, then we know that $k_1+k_2=k$, but --- without more information --- we cannot determine $k_{1,2}$ individually.
\item When a manifest $U(1)$ factor is enhanced to some nonabelian $G_k$, the freedom to change the normalization of the $U(1)$ generator prevents us, in most cases, from being able to compute the level $k$.
\end{itemize}

Often (indeed, in the overwhelming majority of cases), S-dualities, or other considerations, can be used to determine these levels. In the $E_8$ theory, only 244 out of the 924 fixtures with enhanced symmetries were not amenable to those techniques. Here, we introduce a new one, supplementing the analysis of \cite{Chacaltana:2018vhp}, thereby determining almost all of the previously unknown levels.

\subsection{Nilpotent Higgsing}
Let $\mathfrak{f}\subset \mathfrak{f}_\mathcal{T}$ be a simple subalgebra of the flavour symmetry of some SCFT, $\mathcal{T}$, of class-S. We will restrict ourselves to the case where $\mathfrak{f}$ is a simple factor in the manifest flavour symmetry associated to a puncture, $O_1$. Turning on a VEV for the moment map ($\hat{B}_1$ operator) which is the highest root of $\mathfrak{f}$ induces an RG flow which, in favourable circumstances \cite{Distler-Martone}, yields in the infrared another class-S SCFT, $\mathcal{T}'$, with the puncture $O_1$ replaced by the puncture $O_2$, where the nilpotent orbit $O_1$ lies in the closure of the orbit $O_2$.  The key observation of \cite{Beem:2019snk}, which we will review in \S\ref{DSreduction}, is that the flow from $\mathcal{T}$ to $\mathcal{T}'$ is implemented on the level of the chiral algebras as Drinfeld-Sokolov reduction.

In the nilpotent Higgsing of the flavour symmetry associated to a puncture, there are two distinct types. They can most succinctly be characterized by their effect on the Coulomb branch. In the first type, turning on the highest root moment map for some simple factor in the flavour symmetry, $\mathfrak{f}_{2l}$, decreases the Coulomb branch dimension by 1, because we lose a Coulomb branch generator of dimension $\Delta=l$.
\begin{subequations}
\begin{equation}\label{reghiggs}
\begin{gathered}
   O_1\xrightarrow{\;\mathfrak{f}_{2l}\;}O_2\; : \qquad
   \Delta\dim_{\mathbb{H}}(\text{Higgs})=\check{h}(\mathfrak{f})-1,\quad 
   \Delta  n_v = 2l-1
\end{gathered}
\end{equation}
In the second type, the Coulomb branch dimension is preserved. The puncture $O_1$ is Higgsed to $O_2$ in the same special piece. When\footnote{$A(O)$ is the equivariant fundamental group of the nilpotent orbit $O$ (see \cite{CollingwoodMcGovern}). $d$ is the Spaltentstein-Barbasch-Vogan operator \cite{Spaltenstein,BarbaschVogan}, an order-reversing map from the set of nilpotent orbits in $\mathfrak{g}$ to the set of nilpotent orbits in ${}^L\mathfrak{g}$. $d^2O$ is the special orbit in the special piece containing $O$. See \cite{Chacaltana:2012zy} for a physics introduction.} $A(d^2O_1)=\mathbb{Z}_2^k$, this results in replacing a Coulomb branch generator of dimension $\Delta =2l$ with a generator of dimension $\Delta=l$. This is a nilpotent Higgsing, in the sense we have been using the term: it is triggered by turning on a VEV for the highest root moment map. More generally, in type $E$ (including the twisted sector of $E_6$, whose punctures are labeled by nilpotent orbits in $F_4$), $A(d^2O_1)$ can be nonabelian ($S_3$, $S_4$ or $S_5$), so that the Sommers-Achar group can contain a $\mathbb{Z}_n$ factor for $n>2$. In that case there is a Higgsing where a Coulomb branch parameter of dimension $\Delta=nl$ is replaced by a Coulomb branch parameter of dimension $\Delta=l$. This is not quite a nilpotent Higgsing, in that the chiral ring relations require us to turn on VEVs for more than just the moment map\footnote{For instance, in the $l=4,n=3$ cases below, there is a $\hat{B}_{3/2}$ operator, transforming in the spin-$3/2$ representation of the $\mathfrak{su}(2)_{26}$, which gets a VEV.}. Nevertheless, in both case, the Higgsing has the effect:
\begin{equation}\label{snhiggs}
\begin{gathered}
   O_1\xrightarrow{\;\mathfrak{sp}(p)_{(n-1)(nl+1)}\;}O_2\; : \qquad
   \Delta\dim_{\mathbb{H}}(\text{Higgs})=p,\quad 
   \Delta  n_v = 2(n-1)l
\end{gathered}
\end{equation}
\end{subequations}
Note that, in both \eqref{reghiggs} and \eqref{snhiggs}, the decrease in Higgs branch dimension is $\check{h}(\mathfrak{f})-1$.

There is a finite list of special pieces with nonabelian equivariant fundamental group (and hence Higgsings of type \eqref{snhiggs} with $n>2$). Let us list them here, labeling explicitly the Higgsings with $n>2$.

For $E_6$, we have
\[
l=4:\qquad 2A_2+A_1{\color{blue}\xrightarrow[{\color{black}(n=3)}]{{\color{black}\;SU(2)_{26}\;}}}A_3+A_1\xrightarrow{\;SU(2)_9\;}D_4(a_1)
\]
For $F_4$ (twisted $E_6$) we have
\[
l=3:\qquad\begin{matrix}
\begin{tikzpicture}
\node (A2A1t) at (0,0) {$A_2+\tilde{A}_1$};
\node (B2) at (2.5,1) {$B_2$};
\node (A2tA1) at (2.5,-1) {$\tilde{A}_2+A_1$};
\node (C3a1) at (5,0) {$C_3(a_1)$};
\node (F4a3) at (8,0) {$F_4(a_3)$};
\path[thick, ->] 
(A2A1t) edge[dashed] (B2)
(A2A1t) edge[blue] node[black,below left=-.125cm and -.125cm] {$\scriptstyle \underset{(n=4)}{SU(2)_{39}}$} (A2tA1)
(A2tA1) edge[blue] node[black,below right=-.125cm and -.125cm] {$\scriptstyle \underset{(n=3)}{SU(2)_{20}}$} (C3a1)
(B2) edge node [above right=-.125cm and -.125cm] {$\scriptstyle SU(2)_{7}$} (C3a1)
(C3a1) edge node[above] {$\scriptstyle SU(2)_7$} (F4a3)
;
\end{tikzpicture}\end{matrix}
\]
For $E_7$, we have
\[
\begin{aligned}
l=4:& \qquad A_5+A_1{\color{blue}\xrightarrow[{\color{black}(n=3)}]{{\color{black}\;SU(2)_{26}\;}}}D_6(a_2)\xrightarrow{\;SU(2)_9\;}E_7(a_5)\\
l=6:&\qquad 2A_2+A_1{\color{blue}\xrightarrow[{\color{black}(n=3)}]{{\color{black}\;SU(2)_{38}\;}}}(A_3+A_1)'\xrightarrow{\;SU(2)_{13}\;}D_4(a_1)
\end{aligned}
\]
And, finally, for $E_8$,
\[
\begin{aligned}
l=4:& \qquad E_6+A_1{\color{blue}\xrightarrow[{\color{black}(n=3)}]{{\color{black}\;SU(2)_{26}\;}}}E_7(a_2)\xrightarrow{\;SU(2)_9\;}E_8(b_5)\\
l=6:&\qquad \begin{matrix}
\begin{tikzpicture}
\node (A4A3) at (0,0) {$A_4+A_3$};
\node (A5A1) at (2.5,1) {$A_5+A_1$};
\node (D5a1A2) at (2.5,-1) {$D_5(a_1)+A_2$};
\node (E6a3A1) at (6.5,1) {$E_6(a_3)+A_1$};
\node (D6a2) at (6.5,-1) {$D_6(a_2)$};
\node (E7a5) at (9,0) {$E_7(a_5)$};
\node (E8a7) at (12,0) {$E_8(a_7)$};
\path[thick, ->] 
(A4A3) edge[dashed] (A5A1)
(A4A3) edge[blue] node[black,below left=-.125cm and -.125cm] {$\scriptstyle \underset{(n=5)}{SU(2)_{124}}$} (D5a1A2)
(A5A1) edge node[above] {$\scriptstyle SU(2)_{13}$} (E6a3A1)
(D5a1A2) edge[dashed] (D6a2)
(A5A1) edge[blue] node[black,below left=-.65cm and .85cm] {$\scriptstyle SU(2)_{38}$} node[black,above right=.3cm and -.95cm] {$\scriptscriptstyle (n=3)$} (D6a2)
(4.4,-.05) edge[-, white, line width=4.5] (4.6,.05)
(D5a1A2) edge[blue] node[black,above left=-.65cm and 1cm] {$\scriptstyle SU(2)_{75}$} node[black,below right=.25cm and -.9cm] {$\scriptscriptstyle (n=4)$}(E6a3A1)
(E6a3A1) edge[blue] node[black,above right=-.125cm and -.125cm] {$\scriptstyle \overset{(n=3)}{SU(2)_{38}}$} (E7a5)
(D6a2) edge node[below right=-.125cm and -.125cm] {$\scriptstyle SU(2)_{13}$} (E7a5)
(E7a5) edge node[above] {$\scriptstyle SU(2)_{13}$} (E8a7) 
;
\end{tikzpicture}\end{matrix}\\
l=10:&\qquad 2A_2+A_1{\color{blue}\xrightarrow[{\color{black}(n=3)}]{{\color{black}\;SU(2)_{62}\;}}}A_3+A_1\xrightarrow{\;SU(2)_{21}\;}D_4(a_1)\\
       &\qquad 2A_2+2A_1{\color{blue}\xrightarrow[{\color{black}(n=3)}]{{\color{black}\;Sp(2)_{62}\;}}}A_3+2A_1\xrightarrow{\;SU(2)_{21}\;}D_4(a_1)+A_1
\end{aligned}
\]

The full diagram of nilpotent Higgsings for $E_8$ is

\begin{equation*}
\scalebox{.615}{
\begin{tikzpicture}
\node (0) at (0,0) {$0$};
\node[below=1cm of   0] (A1) {$A_1$};
\node[below=1cm of  A1] (2A1)  {$2A_1$};
\node[below=1cm of 2A1] (3A1)  {$3A_1$};
\node[below left=1cm and 1cm of 3A1] (A2)  {$A_2$};
\node[below right=1cm and 1cm of 3A1] (4A1)  {$4A_1$};
\node[below=3cm of 3A1] (A2A1)  {$A_2+A_1$};
\node[below=1cm of A2A1] (A2A1A1)  {${\color{red}A_2+2A_1}$};
\node[below right=1cm and 1cm of A2A1A1] (A2A1A1A1)  {${\color{red}A_2+3A_1}$};
\node[below=1cm of A2A1A1A1] (A2A2)  {$2A_2$};
\node[left=4cm of A2A2] (A3) {$A_3$};
\node[below=1cm of A2A2] (A2A2A1)  {${2A_2+A_1}$};
\node[below=1cm of A2A2A1] (A2A2A1A1)  {${2A_2+2A_1}$};
\node[below=2.5cm of A3] (A3A1) {$A_3+A_1$};
\node[below=1cm of A2A2A1A1] (A3A1A1)  {${\color{red}A_3+2A_1}$};
\node[below=1cm of A3A1A1] (D4a1A1)  {$D_4(a_1)+A_1$};
\node[below=1cm of D4a1A1] (A3A2)  {$A_3+A_2$};
\node[below=1cm of A3A2] (A3A2A1)  {${\color{red}A_3+A_2+A_1}$};
\node[below=1.25cm of A3A2A1] (D4a1A2) {${\color{red}D_4(a_1)+A_2}$};
\node[below=1.25cm of D4a1A2] (A3A3) {$2A_3$};
\node[left=1cm of A3A2A1] (A4)  {$A_4$};
\node[below=3.25cm of A4] (A4A1) {$A_4+A_1$};
\node[below=1.25cm of A4A1] (A4A1A1) {$A_4+2A_1$};
\node[below=1.1cm of A4A1A1] (A4A2dummy) {};
\node[right=15cm of 0] (A4A1A1dummy) {};
\node[below=1cm of A4A1A1dummy] (A4A2) {$A_4+A_2$};
\node[below=1cm of A3A1] (D4a1) {${D_4(a_1)}$};
\node[below=3cm of D4a1] (D4) {$D_4$};
\node[below=3cm of D4] (D4A1) {${D_4+A_1}$};
\node[below=2.75cm of D4A1] (D5a1) {$D_5(a_1)$};
\node[below=1.25cm of D5a1] (D5a1A1dummy) {};
\node[left=2cm of A4A1A1dummy] (D5a1dummy) {};
\node[below left=.75cm and .125cm of A4A2] (D5a1A1) {${\color{red}D_5(a_1)+A_1}$};
\node[right=4cm of D5a1A1] (A4A2A1) {${\color{red}A_4+A_2+A_1}$};
\node[below=2cm of A4A2] (A5) {$A_5$};
\node[below=1cm of A4A2A1] (A4A3) {${A_4+A_3}$};
\node[right=1cm of A4A3] (D4A2) {$D_4+A_2$};
\node[below=2.7cm of D5a1A1] (E6a3) {$E_6(a_3)$};
\node[below=1cm of A5] (A5A1) {$A_5+A_1$};
\node[below=1cm of A4A3] (D5a1A2) {${D_5(a_1)+A_2}$};
\node[below=1cm of A5A1] (E6a3A1) {$E_6(a_3)+A_1$};
\node[below=1cm of D5a1A2] (D6a2) {$D_6(a_2)$};
\node[below=2.7cm of E6a3] (D5) {$D_5$};
\node[below=1cm of E6a3A1] (E7a5) {$E_7(a_5)$};
\node[below=1cm of D5] (D5A1) {$D_5+A_1$};
\node[below=1cm of E7a5] (E8a7) {$E_8(a_7)$};
\node[below=1cm of D5A1] (D6a1) {${D_6(a_1)}$};
\node[below=4.3cm of D6a2] (A6) {$A_6$};
\node[below=1cm of D6a1] (E7a4) {$E_7(a_4)$};
\node[below=1cm of A6] (A6A1) {$A_6+A_1$};
\node[below=1cm of E7a4] (E6a1) {$E_6(a_1)$};
\node[below=1cm of A6A1] (D5A2) {$D_5+A_2$};
\node[below=1cm of D5A2] (D7a2) {$D_7(a_2)$};
\node[below=2.7cm of E6a1] (E6) {$E_6$};
\node[below=7.7cm of E8a7] (A7) {$A_7$};
\node[below=1cm of D7a2] (E6a1A1) {$E_6(a_1)+A_1$};
\node[below=14.5cm of D4A2] (D6) {$D_6$};
\node[below=1cm of A7] (E8b6) {$E_8(b_6)$};
\node[below=1cm of E6a1A1] (E7a3) {$E_7(a_3)$};
\node[below=2.7cm of E6] (E6A1) {${E_6+A_1}$};
\node[below=1cm of E7a3] (D7a1) {$D_7(a_1)$};
\node[below=1cm of E6A1] (E7a2) {$E_7(a_2)$};
\node[below=1cm of D7a1] (E8a6) {$E_8(a_6)$};
\node[below=1cm of E7a2] (E8b5) {$E_8(b_5)$};
\node[below=1cm of E8a6] (D7) {$D_7$};
\node[below=1cm of E8b5] (E7a1) {$E_7(a_1)$};
\node[below=1cm of D7] (E8a5) {$E_8(a_5)$};
\node[below=7.3cm of E8b6] (E8b4) {$E_8(b_4)$};
\node[below=1.5cm of E7a1] (E7) {$E_7$};
\node[below=1.5cm of E8a5] (E8a4) {$E_8(a_4)$};
\node[below=2cm of E8b4] (E8a3) {$E_8(a_3)$};
\node[below=.5cm of E8a3] (E8a2) {$E_8(a_2)$};
\node[below=.5cm of E8a2] (E8a1) {$E_8(a_1)$};
\path[->] (0) edge node[left] {$ (E_8)_{60}$}  (A1)
     (A1) edge node[left] {$ (E_7)_{48}$} (2A1)
    (2A1) edge node[left] {$Spin(13)_{40}$} (3A1)
    (3A1) edge node[above left=-.125cm and -.125cm] {$ SU(2)_{31}$} (A2)
    (3A1) edge node[above right=-.125cm and -.125cm] {$ (F_4)_{36}$} (4A1)
     (A2) edge node[below left=-.125cm and -.125cm] {$ (E_6)_{36}$} (A2A1)
    (4A1) edge node[below right=-.125cm and -.125cm] {$ Sp(4)_{31}$} (A2A1)
   (A2A1) edge node[left] {$ SU(6)_{30}$} (A2A1A1)
 (A2A1A1) edge node[above right=-.125cm and -.125cm] {$ Spin(7)_{28}$} (A2A1A1A1)
(A2A1A1A1) edge node[right] {$ SU(2)_{25}$} (A2A2)
   (A2A2) edge node[right] {$(G_2)_{24}$} (A2A2A1)
 (A2A1A1) edge[dashed] (A3)
     (A3) edge node[left] {$Spin(11)_{28}$} (A3A1)
 (A2A2A1) edge[blue] node[above left=-.125cm and -.125cm,black] {$SU(2)_{62}$} (A3A1)
 (A2A2A1) edge node[right] {$(G_2)_{24}$} (A2A2A1A1)
(A2A2A1A1) edge[blue] node[right,black] {$Sp(2)_{62}$}(A3A1A1)
   (A3A1) edge node[above right=-.125cm and -.125cm] {$Spin(7)_{24}$} (A3A1A1)
   (A3A1) edge node[left] {$SU(2)_{21}$} (D4a1)
 (A3A1A1) edge node[right] {$Sp(2)_{21}$} (D4a1A1)
 (D4a1A1) edge node[right] {$SU(2)_{20}$} (A3A2)
   (D4a1) edge[dashed] (D4)
   (D4a1) edge node[above right=-.125cm and -.125cm] {$Spin(8)_{24}$} (D4a1A1)
   (A3A2) edge node[right] {$Sp(2)_{20}$} (A3A2A1)
   (A3A2) edge[dashed] (A4)
     (D4) edge node[left] {$(F_4)_{24}$} (D4A1)
 (A3A2A1) edge[dashed] (D4A1)
 (A3A2A1) edge node[right] {$SU(2)_{19}$} (D4a1A2)
 (D4a1A2) edge[dashed] (A4A1)
     (A4) edge node[above right=.025cm] {$SU(5)_{20}$}(A4A1)
 (D4a1A2) edge[dashed] (A3A3)
   (A4A1) edge node[above right=0cm and 0cm] {$SU(3)_{18}$}(A4A1A1)
   (A4A1) edge[dashed] (D5a1)
  (D4A1) edge node[left] {$Sp(3)_{19}$}(D5a1)
  (A3A3) edge node[below right=-.125cm and -.125cm] {$Sp(2)_{31}$}(A4A1A1)
  (A4A1A1) edge[-, dash pattern=on .8cm off 1pt on 1pt  off 1pt on 1pt  off 1pt on 1pt  off 1pt on 1pt ] node[right] {$SU(2)_{30}$}(A4A2dummy)
  (A4A1A1dummy) edge[dash pattern=on 1pt  off 1pt on 1pt  off 1pt on 1pt  off 1pt on 1pt  off 1pt on 2cm] node[right] {$SU(2)_{30}$}(A4A2)
  (D5a1) edge[-, dash pattern=on .9cm off 1pt on 1pt  off 1pt on 1pt  off 1pt on 1pt  off 1pt on 1pt ] node[left] {$SU(4)_{18}$} (D5a1A1dummy)
    (D5a1dummy) edge[dash pattern=on 1pt  off 1pt on 1pt  off 1pt on 1pt  off 1pt on 1pt  off 1pt on 2.1cm] node[left] {$SU(4)_{18}$} (D5a1A1)
  (A4A2) edge[dashed] (D5a1A1)
  (A4A2) edge[dashed] (A5)
  (A4A2) edge node[above right=-.125cm and -.125cm] {$SU(2)_{16}$} (A4A2A1)
(D5a1A1) edge[dashed] (E6a3)
(D5a1A1) edge node[above right=.125cm and -2cm] {$SU(2)_{16}$} (D4A2)
(A4A2A1) edge[dashed] (A4A3)
(A4A2A1) edge[dashed] (D4A2)
(A5)edge node[above left=.125cm and -.5cm] {$SU(2)_{13}$} (E6a3)
(A5)edge node[right] {$(G_2)_{16}$} (A5A1)
(A4A3) edge[dashed] (A5A1)
(A4A3) edge[blue] node[left,black] {$SU(2)_{124}$} (D5a1A2)
(D4A2)edge node[below right=-.125cm and -.125cm] {$SU(3)_{28}$} (D5a1A2)
(E6a3) edge[dashed] (D5)
(E6a3)edge node[above right=.125cm and -.5cm] {$(G_2)_{16}$} (E6a3A1)
(A5A1)edge node[right=-.125cm] {$SU(2)_{13}$} (E6a3A1)
(A5A1) edge[blue] node[above right=.375cm and -1cm,black] {$SU(2)_{38}$} (D6a2)
(D5a1A2) edge[blue] node[below right=.125cm and -.75cm,black] {$SU(2)_{75}$} (E6a3A1)
(D5a1A2) edge[dashed] (D6a2)
(E6a3A1) edge[blue] node[left,black] {$SU(2)_{38}$} (E7a5)
(D6a2)edge node[below right=-.125cm and -.125cm] {$SU(2)_{13}$} (E7a5)
(D5) edge node[left] {$Spin(7)_{16}$} (D5A1)
(E7a5) edge[dashed] (D5A1)
(E7a5)edge node[right] {$SU(2)_{13}$} (E8a7)
(D5A1)edge node[left] {$SU(2)_{13}$} (D6a1)
(E8a7)edge[dashed](D6a1)
(E8a7)edge[dashed](A6)
(D6a1)edge node[left] {$SU(2)_{12}$} (E7a4)
(A6)edge node[above left=-.125cm and -.125cm] {$SU(2)_{60}$} (E7a4)
(A6)edge node[right] {$SU(2)_{12}$} (A6A1)
(E7a4)edge[dashed] (E6a1)
(E7a4)edge node[above right=-.125cm and -.125cm] {$SU(2)_{12}$}  (D5A2)
(A6A1)edge node[right] {$SU(2)_{60}$}(D5A2)
(E6a1) edge[dashed](E6)
(E6a1) edge node[above right=-.125cm and -.125cm] {$SU(3)_{12}$}(E6a1A1)
(D5A2) edge[dashed] (D7a2)
(D5A2) edge[dashed] (D6)
(D7a2) edge[dashed] (A7)
(D7a2) edge[dashed] (E6a1A1)
(E6) edge node[left] {$(G_2)_{12}$} (E6A1)
(A7) edge node[left] {$SU(2)_{31}$} (E8b6)
(E6a1A1)edge[dashed] (E8b6)
(E6a1A1)edge[dashed] (E7a3)
(D6) edge node[below right=-.125cm and -.125cm] {$Sp(2)_{11}$} (E7a3)
(E8b6) edge[dashed] (E6A1)
(E8b6) edge[dashed] (D7a1)
(E7a3)edge node[right] {$SU(2)_{10}$}(D7a1)
(E6A1)edge[blue] node[left,black] {$SU(2)_{26}$}(E7a2)
(D7a1)edge[dashed](E7a2)
(D7a1)edge[dashed](E8a6)
(E7a2)edge node[left] {$SU(2)_{9}$}(E8b5)
(E8a6)edge[dashed](E8b5)
(E8a6)edge[dashed](D7)
(E8b5)edge[dashed](E7a1)
(E8b5)edge[dashed](E8a5)
(D7)edge node[right] {$SU(2)_{13}$}(E8a5)
(E7a1)edge node[below left=-.125cm and -.125cm]{$SU(2)_{8}$}(E8b4)
(E8a5)edge[dashed](E8b4)
(E8b4)edge[dashed](E7)
(E8b4)edge[dashed](E8a4)
(E7)edge node[below left=-.125cm and -.125cm]{$SU(2)_{7}$}(E8a3)
(E8a4)edge[dashed](E8a3)
(E8a3)edge[dashed](E8a2)
(E8a2)edge[dashed](E8a1)
;
\end{tikzpicture}
}
\end{equation*}

This is a decorated version of the Hasse diagram for nilpotent orbit closure.  Solid lines indicate nilpotent Higgsings, while dashed lines indicate nilpotent orbit closures with no corresponding nilpotent Higgsing. The blue lines represent Higgsings of type \eqref{snhiggs} with $n>2$. Nilpotent orbits marked in red have simple flavour symmetry factors whose possible Higgsing do not correspond to nilpotent orbit closure. 

For \eqref{reghiggs} and for \eqref{snhiggs} when $n=2$, the nilpotent Higgsings we described have an elegant interpretation, in terms of the 2D Vertex Operator Algebra associated to the 4D $\mathcal{N}=2$ SCFT, which we review in the next subsection

\subsection{Drinfeld-Sokolov Reduction}\label{DSreduction}

After restricting to a plane in $\mathbb{R}^4$, and applying a suitable twisted notion of translations in the plane, the Schur operators of a 4D $\mathcal{N}=2$ SCFT form a 2D chiral algebra \cite{Beem:2013sza}. The moment map ($\hat{B}_1$) operators which generate the flavour symmetry algebra become the currents of a 2D current algebra, whose level
\begin{equation}\label{k2d4d}
k_{2D}= -\tfrac{1}{2} k_{4D}
\end{equation}
Similarly, the Virasoro central charge of the 2D theory is related to the Weyl anomaly coefficient of the 4D theory by
\begin{equation}\label{c2d4d}
c_{2D}= -12 c_{4D}
\end{equation}
Now, say we want to turn on a VEV for the highest root moment map of some simple factor $\mathfrak{f}_k$ in the 4D theory. This has an elegant description in terms of the 2D chiral algebra \cite{Beem:2014rza,Beem:2019tfp,Beem:2019snk}.

Let $\lambda$ be the highest root of $\mathfrak{f}$ and let $\beta_a$, $a=1,2,\dots,\check{h}(\mathfrak{f})-2$ be a collection of positive roots with the property that
\begin{subequations}\label{betadef}
\begin{align}
\lambda\cdot\beta_a&=1\label{betadefa}\\
\beta_a+\beta_{a'}&\;\text{is not a root}\label{betadefb}
\end{align}
\end{subequations}
Up to some obvious automorphisms of the algebra, the choice of $\{\beta_a\}$ is unique, and we list them in Appendix \ref{App:Pos}.

Let $J_\lambda(z)$ and $J_{\beta_a}(z)$ be the corresponding 2D currents and let $H_i(z)$ be the currents corresponding to the Cartan generators. Because of \eqref{betadefb}, the only singular OPEs between these currents are
\begin{equation}
\begin{split}
H_i(z)H_j(w)&= \frac{k_{2D} \delta_{ij}}{(z-w)^2}\\
H_i(z) J_{\beta_a}(w)&= \frac{\beta_{ai}J_{\beta_a}(w)}{z-w}
\end{split}
\end{equation}
Introduce a set of $(\check{h}(\mathfrak{f})-2)$ spin-1/2 $bc$ systems and one spin-1 $bc$ system. If $T(z)$ is the stress tensor of the original VOA, construct the ``improved" stress tensor
\begin{equation}\label{improved}
T_{\text{new}}= T- \tfrac{1}{2}\lambda\cdot \partial H +(\partial b_\lambda)c^\lambda +\sum_{a=1}^{\check{h}(\mathfrak{f})-2} \tfrac{1}{2} (\partial b_a)c^a -\tfrac{1}{2}b_a(\partial c^a)
\end{equation}
With this improved stress tensor, the conformal weights $h(J_\lambda)=0$ and\footnote{Overall, $2\check{h}(f)-4$ currents corresponding to positive roots have their conformal weights shifted to $h=1/2$. We need to choose a collection of half of them which together satisfy \eqref{betadefb}. Given a positive root $\beta_a$ satisfying \eqref{betadef}, then $\gamma_a=\lambda-\beta_a$ is another positive root which satisfies $\lambda\cdot\gamma_a=1$. The corresponding currents satisfy
\[
J_{\beta_a}(z) J_{\gamma_b}(w) =\frac{\delta_{ab} J_\lambda}{z-w}
\]
Because our BRST fixing sets $J_\lambda$ to a nonzero constant, the $J_{\beta_a}$ are a maximal set of spin-1/2 currents that we can set to zero. In the BRST cohomology, both sets of currents drop out: $J_{\beta_a}(z)= \{Q,b_a(z)\}$ and $[Q,J_{\gamma_a}(z)]\sim e c_a(z)$.} $h(J_{\beta_a})=1/2$. We now define a nilpotent BRST operator
\begin{equation}
Q= \oint \frac{dz}{2\pi i} \Bigl[(J_\lambda(z)-e)c^\lambda(z) +\sum_{a=1}^{\check{h}(f)-2}J_{\beta_a}(z)c^a(z)\Bigr]
\end{equation}
where $e$ is some nonzero complex constant. It is evident that $Q^2=0$ and $[Q,T_{\text{new}}(z)]=0$. The claim of \cite{Beem:2014rza,Beem:2019snk} is that the VOA of the IR SCFT is the $Q$-cohomology of the (improved) VOA  of the UV theory.

This has some immediate consequences. The Virasoro central charge of the new VOA is read off from the most singular term of the OPE of $T_{\text{new}}$ with itself.
\begin{equation}
\begin{split}
c^{\text{new}}&= c -3k |\lambda|^2 -2 +(\check{h}(\mathfrak{f})-2)\\
&= c - 6k + \check{h}(\mathfrak{f}) -4
\end{split}
\end{equation}
Using \eqref{k2d4d},\eqref{c2d4d}, the $c$ Weyl anomaly coefficient of the 4D theory is
\begin{equation}\label{c4dchange}
12 c^{\text{new}}_{4D}= 12 c_{4D} -3 (k_{4D} -1)- (\check{h}(\mathfrak{f})-1)
\end{equation}
In turn, the Weyl anomaly coefficient is related to the effective number of vectors and hypers in the 4d theory by
\begin{equation}\label{c4drel}
\begin{split}
12 c_{4D}&= 2n_v+n_h\\
& = 3n_v +(n_h-n_v)\\
&= 3 n_v +\dim_\mathbb{H}\text{Higgs}
\end{split}
\end{equation}
where we used that we are in the situation where there is no unbroken gauge symmetry on the Higgs branch. Taking the BRST cohomology killed $(\check{h}(\mathfrak{f})-1)$ generators, so the quaternionic dimension of the Higgs branch decreased by $\check{h}(\mathfrak{f})-1$. Combining this with \eqref{c4dchange} and \eqref{c4drel}, we find the decrease in the effective number of vector multiplets is
\begin{equation}
\delta n_v = k_{4D}-1
\end{equation}
in accord with \eqref{reghiggs} and with \eqref{snhiggs} for $n=2$. 

For $n>2$, there's no simple realization of the Higgsing as Drinfeld-Sokolov reduction. Nevertheless, there is such a realization if we are willing to back up one step in the Hasse diagram and Higgs down using a non-highest-root moment map. We'll defer a complete treatment to a followup paper\cite{DistlerElliotWIP}, and here just illustrate the construction for the cases we need for our $E_8$ analysis.

Consider
\[
\begin{tikzpicture}
\node (E6) at (0,0) {$E_6$};
\node (E6A1) at (2,0) {$E_6+A_1$};
\node (E7a2) at (5,0) {$E_7(a_2)$};
\node (E8b5) at (7.5,0) {$E_8(b_5)$};
\path[thick, ->,draw,color=blue] 
(E6) edge[color=black] node[below] {$\scriptstyle (G_2)_{12}$} (E6A1)
(E6A1) edge node[below] {$\scriptstyle SU(2)_{26}$} (E7a2)
(E7a2) edge[color=black] node[below] {$\scriptstyle SU(2)_{9}$} (E8b5)
(0.2,0.2) arc[dashed,radius=4.2, start angle=120, end angle=60,red,draw] node[above right=0.5cm and -2.75cm] {$\scriptstyle (G_2)_{12}$} (E7a2)
;
\end{tikzpicture}
\]
The Higgsing from $E_6$ to $E_6+A_1$ is the conventional one, triggered by turning on a VEV for the highest root moment map of $G_2$. The DS reduction is just as we have described. Let $\lambda=3\alpha_1+2\alpha_2$ be the highest root and $\beta_1=3\alpha_1+\alpha_2$, $\beta_2=2\alpha_1+\alpha_2$ and $\gamma_a=\lambda-\beta_a$. The stress tensor for the VOA where the $E_6$ puncture is replaced by $E_6+A_1$ is
\[
T_{E_6+A_1}= T_{E_6} - \tfrac{1}{2}\lambda\cdot \partial H +(\partial b_1)c_1 +\tfrac{1}{2}(\partial b_2)c_2-\tfrac{1}{2} b_2\partial c_2
+\tfrac{1}{2}(\partial b_3)c_3-\tfrac{1}{2} b_3\partial c_3
\]
on which we take the BRST cohomology of
\[
Q=\oint \frac{dz}{2\pi i} \Bigl[(J_\lambda(z) -e_1) c_1(z) +J_{\beta_1}(z) c_2(z) + J_{\beta_2}(z) c_3(z) \Bigr]
\]
Here, we introduced one spin-1 $bc$ system and two spin-$1/2$ $bc$ systems; $\delta\dim_\mathbb{H}\text{Higgs}=3$ and $\delta n_v = k-1=11$, as we expect when losing one Coulomb branch parameter with $\Delta=6$.

The Higgsing from $E_6$ to $E_7(a_2)$ is instead triggered by turning on a VEV for the $\beta_2$ moment map. The stress tensor is
\begin{equation}\label{TG2nonmin}
T_{E_7(a_2)}=T_{E_6} - \tfrac{3}{2}\beta_2\cdot \partial H +\tfrac{3}{2}(\partial b_1)c_1 +\tfrac{1}{2}b_1\partial c_1
 +\tfrac{3}{2}(\partial b_2)c_2 +\tfrac{1}{2}b_2\partial c_2 +(\partial b_3)c_3
 + \tfrac{1}{2}(\partial b_4)c_4 - \tfrac{1}{2}b_4\partial c_4
\end{equation}
where we introduced two spin-3/2 $bc$ systems, one spin-1 $bc$ system and one spin-1/2 $bc$ system. The BRST operator is
\begin{equation}\label{QG2nonmin}
Q= \oint\frac{dz}{2\pi i} \Bigl[J_\lambda(z) c_1(z) +J_{\beta_1}(z) c_2(z) + (J_{\beta_2}(z)-e_2) c_3(z) +J_{\gamma_2}(z) c_4(z)  - b_1(z)c_3(z) c_4(z)\Bigr]
\end{equation}
The Virasoro central charge
\[\begin{split}
c_{E_7(a_2)}&=c_{E_6} -12 \left(\tfrac{3}{2}\right)^2|\beta_2|^2 k_{2D}  +2(-11) +(-2) +(1)\\
&=c_{E_6} -18 k_{2D} -23
\end{split}
\]
As before, we relate these to the 4D quantities, using \eqref{k2d4d}, \eqref{c2d4d} and  \eqref{c4drel}, which yield
\begin{equation}
3\delta n_v +\delta\dim_\mathbb{H}\text{Higgs} = 9k-23
\end{equation}
From \eqref{QG2nonmin}, we have that the decrease in the Higgs branch dimension $\delta\dim_\mathbb{H}\text{Higgs}=4$. Thus, since $k=12$, we have
\begin{equation*}
\delta n_v =3k-9=27
\end{equation*}
which is the correct answer when we lose one Coulomb branch generator with $\Delta=6$ and replace another Coulomb branch generator having $\Delta=12$ with one having $\Delta=3$.

Finally, we will have recourse to a Higgsing which turns on a moment map in the regular nilpotent orbit of $\mathfrak{su}(3)$. As before, let $\lambda=\alpha_1+\alpha_2$ be the highest root, and consider the ``improved" stress tensor
\begin{equation}
T_{\text{new}}= T-\lambda\cdot \partial H +2(\partial b_\lambda)c^\lambda+b_\lambda\partial c^\lambda+(\partial b_1)c^1+ (\partial b_2)c^2
\end{equation}
(Notice the factor of 2 relative to \eqref{improved}.) With this stress tensor, $h(J_\lambda)=-1$ and $h(J_{\alpha_1})=h(J_{\alpha_2})=0$. So the BRST operator in this case
\begin{equation}
Q=\oint\frac{dz}{2\pi i} \Bigl[J_\lambda(z)c^\lambda(z) +(J_{\alpha_1}(z)-e_1)c^1(z)+(J_{\alpha_2}(z)-e_2)c^2(z)-b_\lambda(z) c^1(z) c^2(z)\Bigr]
\end{equation}
has two c-number parameters, $e_1$,$e_2$.

Applying the same relations between the 2D and 4D quantities as before, this gives
\begin{equation}
\begin{gathered}
\delta n_v= 4k_{4D}-11\\
\delta\dim\text{Higgs}= 3
\end{gathered}
\end{equation}
We'll use this for 
\[
\begin{tikzpicture}
\node (E6a1) at (0,0) {$E_6(a_1)$};
\node (E6a1A1) at (3.5,0) {$E_6(a_1)+A_1$};
\node (E8b6) at (6,0) {$E_8(b_6)$};
\path[thick, ->,draw,color=blue] 
(E6a1) edge[color=black] node[below] {$\scriptstyle SU(3)_{12}$} (E6a1A1)
(E6a1A1) edge[black,dashed]  (E8b6)
(0.6,0.2) arc[dashed,radius=4.8, start angle=120, end angle=60,red,draw] node[above right=0.6cm and -3cm] {$\scriptstyle SU(3)_{12}$} (E8b6)
;
\end{tikzpicture}
\]

\section{The \texorpdfstring{$E_8$}{E₈} Theory}\label{the_E8_theory}
\subsection{Results}
Of the 49,836 fixtures in the $E_8$ theory,  924 have enhanced global symmetry and/or free hypermultiplets. Of these 924, 244 had undetermined levels. We fix 235 of those (see Table \ref{fixedfixtures}), leaving the levels in 9 fixtures undetermined (see Table \ref{unfixedfixtures}). We also fix 23 typos in the \href{https://golem.ph.utexas.edu/class-S/E8/}{application} (these are marked in {\color{red}red} in table \ref{fixedfixtures}).

{
\footnotesize
\setlength\LTleft{-.25in}
\renewcommand{\arraystretch}{2.25}

\begin{longtable}{|c|c|c|c|}
\caption{Corrected Fixtures}\label{fixedfixtures}\\
\hline
\#&Fixture&Old Flavour Symmetry&New Flavour Symmetry\\
\endfirsthead
\hline
\#&Fixture&Old Flavour Symmetry&New Flavour Symmetry\\
\endhead
\endfoot
\hline
1&$A_6,\;E_6,\;E_6$&$\begin{gathered}{SU(2)}_{12}\times {SU(2)}_{60-k}\\ \times {SU(2)}_k\times {(G_2)}_{12}^2\end{gathered}$&$\begin{gathered}{SU(2)}_{12}\times {SU(2)}_{48}\\ \times {SU(2)}_{12}\times {(G_2)}_{12}^2\end{gathered}$\\
\hline
2&$A_6,\;E_6,\;E_6+A_1$&$\begin{gathered}{SU(2)}_{12}\times {SU(2)}_{60-k}\\ \times {SU(2)}_k\times {(G_2)}_{12}\\ \times {SU(2)}_{26}\end{gathered}$&$\begin{gathered}{SU(2)}_{12}\times {SU(2)}_{48}\\ \times {SU(2)}_{12}\times {(G_2)}_{12}\\ \times {SU(2)}_{26}\end{gathered}$\\
\hline
3&$A_6+A_1,\;E_6,\;E_6+A_1$&$\begin{gathered}{SU(2)}_{60-k}\times {SU(2)}_k\\ \times {(G_2)}_{12}\times {SU(2)}_{26}\end{gathered}$&$\begin{gathered}{SU(2)}_{48}\times {SU(2)}_{12}\\ \times {(G_2)}_{12}\times {SU(2)}_{26}\end{gathered}$\\
\hline
4&$A_6,\;E_6+A_1,\;E_6+A_1$&$\begin{gathered}{SU(2)}_{12}\times {SU(2)}_{60-k}\\ \times {SU(2)}_k\times {SU(2)}_{26}^2\end{gathered}$&$\begin{gathered}{SU(2)}_{12}\times {SU(2)}_{48}\\ \times {SU(2)}_{12}\times {SU(2)}_{26}^2\end{gathered}$\\
\hline
5&$A_6+A_1,\;E_6+A_1,\;E_6+A_1$&$\begin{gathered}{SU(2)}_{60-k}\times {SU(2)}_k\\ \times {SU(2)}_{26}^2\end{gathered}$&$\begin{gathered}{SU(2)}_{48}\times {SU(2)}_{12}\\ \times {SU(2)}_{26}^2\end{gathered}$\\
\hline
6&$A_6,\;E_6,\;E_7(a_2)$&$\begin{gathered}{SU(2)}_{12}\times {SU(2)}_{60-k}\\ \times {SU(2)}_k\times {(G_2)}_{12}\\ \times {SU(2)}_9\end{gathered}$&$\begin{gathered}{SU(2)}_{12}\times {SU(2)}_{48}\\ \times {SU(2)}_{12}\times {(G_2)}_{12}\\ \times {SU(2)}_9\end{gathered}$\\
\hline
7&$A_6+A_1,\;E_6,\;E_7(a_2)$&$\begin{gathered}{SU(2)}_{60-k}\times {SU(2)}_k\\ \times {(G_2)}_{12}\times {SU(2)}_9\end{gathered}$&$\begin{gathered}{SU(2)}_{48}\times {SU(2)}_{12}\\ \times {(G_2)}_{12}\times {SU(2)}_9\end{gathered}$\\
\hline
\end{longtable}

}

%
%

There remain 9 fixtures, whose current algebra levels we were still not able to determine.

{
\footnotesize
\renewcommand{\arraystretch}{2.26}

\begin{longtable}{|c|c|c|}
\caption{Fixtures with unknown levels}\label{unfixedfixtures}\\
\hline
\#&Fixture&Flavour Symmetry\\
\endfirsthead
\hline
\#&Fixture&Flavour Symmetry\\
\endhead
\endfoot
\hline
1&$A_3+A_2,\;D_7,\;E_8(a_4)$&$\begin{gathered}{Sp(2)}_{20}\times {U(1)}^2\\ \times {SU(2)}_{13}\times {SU(2)}_k\end{gathered}$\\
\hline
2&$A_3+A_2+A_1,\;D_7,\;E_8(a_4)$&$\begin{gathered}{SU(2)}_{384-4k}\times {SU(3)}_{k}\\ \times {SU(2)}_{19}\times {SU(2)}_{13}\end{gathered}$\\
\hline
3&$D_4(a_1)+A_2,\;D_7,\;E_8(a_4)$&$\begin{gathered}{SU(3)}_{k_1}\times {SU(3)}_{k_2}\\ \times {SU(3)}_{96-k_1-k_2}\times {SU(2)}_{13}\end{gathered}$\\
\hline
4&$A_3+A_2,\;E_8(a_5),\;E_8(a_4)$&$\begin{gathered}{Sp(2)}_{20}\times {U(1)}^2\\ \times {SU(2)}_{k_1}\times {SU(2)}_{k_2}\end{gathered}$\\
\hline
5&$A_3+A_2+A_1,\;E_8(a_5),\;E_8(a_4)$&$\begin{gathered}{SU(3)}_{96-k}\times {SU(3)}_k\\ \times {SU(2)}_{19}\end{gathered}$\\
\hline
6&$D_4(a_1)+A_2,\;E_8(a_5),\;E_8(a_4)$&$\begin{gathered}{SU(3)}_{k_1}\times {SU(3)}_{k_2}\\ \times {SU(3)}_{k_3}\times {SU(3)}_{96-k_1-k_2-k_3}\end{gathered}$\\
\hline
7&$A_4+A_3,\;D_6(a_2),\;E_8(a_2)$&$\begin{gathered}{SU(2)}_{96-k_1-k_2}\times {SU(2)}_{k_1}\\ \times {SU(2)}_{k_2}\times {Sp(2)}_{13}^2\end{gathered}$\\
\hline
8&$2A_2+A_1,\;D_7(a_2),\;E_8(a_2)$&$\begin{gathered}{SO(7)}_{24}\times {SU(3)}_{k}\\ \times {SU(2)}_{61-k}\times U(1)\end{gathered}$\\
\hline
9&$2A_2+2A_1,\;D_7(a_2),\;E_8(a_2)$&$\begin{gathered}{SU(5)}_k\times {Sp(2)}_{61-k}\end{gathered}$\\
\hline
\end{longtable}
}

\subsection{Methods}

We primarily used nilpotent Higgsing along with S-duality to arrive at our results, see \cite{Distler:2022nsn} for examples. Additionally we matched with other class-S theories to determine the levels of two fixtures.

Consider the pair 
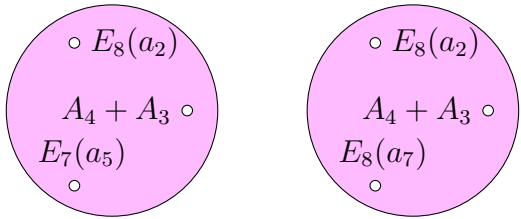
\begin{equation*}
%
\end{equation*}
The interacting part of the fixture on the left was determined to have flavour symmetry ${SU(2)}^6 \times  {Sp(2)_{13}}$ with $SU(2)$ levels adding up to 108 and the interacting part of the fixture on the right has flavour symmetry $SU(2)^{12}$ with the levels adding up to 120. A nilpotent Higgsing of an $SU(2)_{13}$ for the fixture on the left results in the fixture on the right.

 Candidates for isomorphic SCFTs are the interacting parts of $E_7$ type fixtures $[E_6, (A_3+A_1)', 2A_2+A_1]$ and $[E_6, D_4(a_1), 2A_2+A_1]$ respectively. These have the same flavour symmetry, central charges, and graded Coulomb branch dimensions. Computing the Schur index of the fixture on the left to order $\tau^4$ and comparing with that of the interacting part of $[E_6, (A_3+A_1)', 2A_2+A_1]$ we find both indices to be

\begin{equation}
1 +28\tau^2 +68\tau^3+619\tau^4+O(\tau^5)
\end{equation}

We believe this to be sufficient evidence that both theories are indeed identical, hence the two theories obtained by a nilpotent Higgsing of the $SU(2)_{13}$ are identical. This allows us to fill in the missing levels for these two fixtures in Table 1.

\section*{Acknowledgements}
We would like to thank Monica Kang and Craig Lawrie for useful discussions, and Mario Martone for the collaboration which inspired us to find applications for this class of nilpotent Higgsings. Some of this work was pursued at the 2022 Aspen Winter Workshop ``Geometrization of (S)QFTs in $D\leq6$". The first author gratefully acknowledges the Aspen Center for Physics, which is supported by the National Science Foundation Grant No.~PHY--1607611. This work was supported in part by the National Science Foundation under Grants No.~PHY--1914679 and PHY--2210562. 

\begin{appendices}
\section{Positive Roots}\label{App:Pos}
In this appendix, we list the (unique up to the $\mathbb{Z}_2$ automorphisms of the $\mathfrak{su}(n)$ and $\mathfrak{e}_6$ Dynkin diagrams) set of positive roots satisfying
\begin{equation}\label{betacondsagain}
\begin{split}
\lambda\cdot\beta_a&=1\\
\beta_a+\beta_{a'}&\;\text{is not a root}
\end{split}
\end{equation}
for $a=1,2,\dots,\check{h}(\mathfrak{f})-2$ in all the simple Lie algebras.

Let $\alpha_1,\alpha_2,\dots,\alpha_{\operatorname{rank}(\mathfrak{f})}$ be the simple roots and let $\lambda$ be the highest root.
\begin{itemize}
\item[$\mathfrak{su}(n)$:]\qquad $\check{h}(\mathfrak{su}(n))-2=n-2$
  \[
   \begin{split}
   \lambda&= \alpha_1+\alpha_2+\dots+\alpha_{n-1}\\
   \beta_1&= \alpha_1\\
   \beta_2&=\alpha_1+\alpha_2\\
   \beta_3&=\alpha_1+\alpha_2+\alpha_3\\
   &\vdots\\
   \beta_{n-3}&=\alpha_1+\alpha_2+\dots+\alpha_{n-3}\\
   \beta_{n-2}&=\alpha_{n-1}
   \end{split}
  \]

\item[$\mathfrak{so}(2n)$:]\qquad $\check{h}(\mathfrak{so}(2n))-2=2n-4$
  \[
   \begin{split}
   \lambda&= \alpha_1+2\alpha_2+\dots+2\alpha_{n-2}+\alpha_{n-1}+\alpha_n\\
   \beta_1&= \alpha_1+\alpha_2\\
   \beta_2&=\alpha_1+\alpha_2+\alpha_3\\
   &\vdots\\
   \beta_{n-3}&=\alpha_1+\alpha_2+\dots+\alpha_{n-2}\\
   \beta_{n-2}&=\alpha_1+\alpha_2+\dots+\alpha_{n-2}+\alpha_{n-1}\\
   \beta_{n-1}&=\alpha_1+\alpha_2+\dots+\alpha_{n-2}+\alpha_{n}\\
  \beta_{n}&=\alpha_1+\alpha_2+\dots+\alpha_{n-2}+\alpha_{n-1}+\alpha_{n}\\
  \beta_{n+1}&=\alpha_1+\alpha_2+\dots+\alpha_{n-3}+2\alpha_{n-2}+\alpha_{n-1}+\alpha_{n}\\
  \beta_{n+2}&=\alpha_1+\alpha_2+\dots+2\alpha_{n-3}+2\alpha_{n-2}+\alpha_{n-1}+\alpha_{n}\\
     &\vdots\\
  \beta_{2n-4}&=\alpha_1+\alpha_2+2\alpha_3+\dots+2\alpha_{n-2}+\alpha_{n-1}+\alpha_{n}\\
   \end{split}
  \]

\item[$\mathfrak{so}(2n+1)$:]\qquad $\check{h}(\mathfrak{so}(2n+1))-2=2n-3$
  \[
   \begin{split}
   \lambda&= \alpha_1+2\alpha_2+\dots+2\alpha_{n-1}+2\alpha_n\\
   \beta_1&= \alpha_1+\alpha_2\\
   \beta_2&=\alpha_1+\alpha_2+\alpha_3\\
   &\vdots\\
   \beta_{n-2}&=\alpha_1+\alpha_2+\dots+\alpha_{n-1}\\
   \beta_{n-1}&=\alpha_1+\alpha_2+\dots+\alpha_{n-1}+\alpha_{n}\\
   \beta_{n}&=\alpha_1+\alpha_2+\dots+\alpha_{n-1}+2\alpha_{n}\\
  \beta_{n+1}&=\alpha_1+\alpha_2+\dots+\alpha_{n-2}+2\alpha_{n-1}+2\alpha_{n}\\
  \beta_{n+2}&=\alpha_1+\alpha_2+\dots+2\alpha_{n-2}+2\alpha_{n-1}+2\alpha_{n}\\
     &\vdots\\
  \beta_{2n-3}&=\alpha_1+\alpha_2+2\alpha_3+\dots+2\alpha_{n-1}+2\alpha_{n}\\
   \end{split}
  \]

\item[$\mathfrak{sp}(n)$:]\qquad $\check{h}(\mathfrak{sp}(n))-2=n-1$
  \[
   \begin{split}
   \lambda&= 2\alpha_1+2\alpha_2+\dots+2\alpha_{n-1}+\alpha_n\\
   \beta_1&= \alpha_1+\alpha_2+\dots+\alpha_n\\
   \beta_2&=\alpha_1+\alpha_2+\dots+\alpha_{n-2}+2\alpha_{n-1}+\alpha_n\\
   \beta_3&=\alpha_1+\alpha_2+\dots+2\alpha_{n-2}+2\alpha_{n-1}+\alpha_n\\
   &\vdots\\
   \beta_{n-1}&=\alpha_1+2\alpha_2+\dots+2\alpha_{n-1}+\alpha_n\\
   \end{split}
  \]

\item[$\mathfrak{g}_2$:]\qquad $\check{h}(\mathfrak{g}_2)-2=2$
  \[
   \begin{split}
   \lambda&= 3\alpha_1+2\alpha_2\\
   \beta_1&= 3\alpha_1+\alpha_2\\
   \beta_2&=2\alpha_1+\alpha_2
   \end{split}
  \]

\item[$\mathfrak{f}_4$:]\qquad $\check{h}(\mathfrak{f}_4)-2=7$
  \[
   \begin{split}
   \lambda&= 2\alpha_1+3\alpha_2+4\alpha_3+2\alpha_4\\
   \beta_1&=\alpha_1+ 3\alpha_2+4\alpha_3+2\alpha_4\\
   \beta_2&=\alpha_1+ 2\alpha_2+4\alpha_3+2\alpha_4\\
   \beta_3&=\alpha_1+ 2\alpha_2+3\alpha_3+2\alpha_4\\
   \beta_4&=\alpha_1+ 2\alpha_2+3\alpha_3+\alpha_4\\
   \beta_5&=\alpha_1+ 2\alpha_2+2\alpha_3+2\alpha_4\\
   \beta_6&=\alpha_1+ 2\alpha_2+2\alpha_3+\alpha_4\\
   \beta_7&=\alpha_1+2\alpha_2+ 2\alpha_3
   \end{split}
  \]

\item[$\mathfrak{e}_6$:]\qquad $\check{h}(\mathfrak{e}_6)-2=10$
  \[
   \begin{split}
   \lambda&= \alpha_1+2\alpha_2+3\alpha_3+2\alpha_4+\alpha_5+2\alpha_6\\
   \beta_1&=\alpha_1+2\alpha_2+3\alpha_3+2\alpha_4+\alpha_5+\alpha_6\\
   \beta_2&=\alpha_1+2\alpha_2+2\alpha_3+2\alpha_4+\alpha_5+\alpha_6\\
   \beta_3&=\alpha_1+2\alpha_2+2\alpha_3+\alpha_4+\alpha_5+\alpha_6\\
   \beta_4&=\alpha_1+ \alpha_2+2\alpha_3+2\alpha_4+\alpha_5+\alpha_6\\
   \beta_5&=\alpha_1+ 2\alpha_2+ 2\alpha_3+\alpha_4+ \alpha_6\\
   \beta_6&=\alpha_1+ \alpha_2+ 2\alpha_3+\alpha_4+ \alpha_5+\alpha_6\\
   \beta_7&=\alpha_1+ \alpha_2+2\alpha_3+\alpha_4+\alpha_6\\
   \beta_8&=\alpha_1+\alpha_2+\alpha_3+\alpha_4+ \alpha_5+ \alpha_6\\
   \beta_9&=\alpha_1+\alpha_2+ \alpha_3+\alpha_4+ \alpha_6\\
   \beta_{10}&=\alpha_1+\alpha_2+\alpha_3+\alpha_6
   \end{split}
  \]

\item[$\mathfrak{e}_7$:]\qquad $\check{h}(\mathfrak{e}_7)-2=16$
  \[
   \begin{split}
   \lambda&=2\alpha_1+3\alpha_2+4\alpha_3+ 3\alpha_4+ 2\alpha_5+\alpha_6+ 2\alpha_7\\
   \beta_1&=\alpha_1+ 3\alpha_2+ 4\alpha_3+3\alpha_4+2\alpha_5+\alpha_6+2\alpha_7\\
   \beta_2&=\alpha_1+ 2\alpha_2+4\alpha_3+3\alpha_4+2\alpha_5+\alpha_6+2\alpha_7\\
   \beta_3&=\alpha_1+2\alpha_2+3\alpha_3+ 3\alpha_4+2\alpha_5+\alpha_6+2\alpha_7\\
   \beta_4&=\alpha_1+ 2\alpha_2+3\alpha_3+2\alpha_4+2\alpha_5+\alpha_6+ 2\alpha_7\\
   \beta_5&=\alpha_1+ 2\alpha_2+3\alpha_3+ 3\alpha_4+2\alpha_5+\alpha_6+ \alpha_7\\
   \beta_6&=\alpha_1+2\alpha_2+3\alpha_3+ 2\alpha_4+\alpha_5+\alpha_6+2\alpha_7\\
   \beta_7&=\alpha_1+ 2\alpha_2+ 3\alpha_3+2\alpha_4+2\alpha_5+\alpha_6+\alpha_7\\
   \beta_8&=\alpha_1+2\alpha_2+3\alpha_3+2\alpha_4+\alpha_5+2\alpha_7\\
   \beta_9&=\alpha_1+ 2\alpha_2+3\alpha_3+2\alpha_4+\alpha_5+\alpha_6+\alpha_7\\
   \beta_{10}&=\alpha_1+ 2\alpha_2+2\alpha_3+ 2\alpha_4+ 2\alpha_5+\alpha_6+\alpha_7\\
   \beta_{11}&=\alpha_1+2\alpha_2+3\alpha_3+ 2\alpha_4+\alpha_5+ \alpha_7\\
   \beta_{12}&=\alpha_1+ 2\alpha_2+2\alpha_3+2\alpha_4+\alpha_5+\alpha_6+\alpha_7\\
   \beta_{13}&=\alpha_1+ 2\alpha_2+2\alpha_3+ 2\alpha_4+\alpha_5+\alpha_7\\
   \beta_{14}&=\alpha_1+2\alpha_2+2\alpha_3+\alpha_4+\alpha_5+\alpha_6+ \alpha_7\\
   \beta_{15}&=\alpha_1+2\alpha_2+ 2\alpha_3+\alpha_4+ \alpha_5+\alpha_7\\
   \beta_{16}&=\alpha_1+2\alpha_2+2\alpha_3+\alpha_4+\alpha_7
   \end{split}
  \]

\item[$\mathfrak{e}_8$:]\qquad $\check{h}(\mathfrak{e}_8)-2=28$
  \[
   \begin{split}
   \lambda&=2\alpha_1+4\alpha_2+ 6\alpha_3+ 5\alpha_4+4\alpha_5+3\alpha_6+2\alpha_7+3\alpha_8\\
   \beta_1&=2\alpha_1+4\alpha_2+6\alpha_3+ 5\alpha_4+ 4\alpha_5+ 3\alpha_6+\alpha_7+3\alpha_8\\
   \beta_2&=2\alpha_1+4\alpha_2+6\alpha_3+ 5\alpha_4+4\alpha_5+ 2\alpha_6+\alpha_7+ 3\alpha_8\\
   \beta_3&=2\alpha_1+4\alpha_2+6\alpha_3+ 5\alpha_4+ 3\alpha_5+2\alpha_6+\alpha_7+ 3\alpha_8\\
   \beta_4&=2\alpha_1+ 4\alpha_2+ 6\alpha_3+  4\alpha_4+ 3\alpha_5+2\alpha_6+\alpha_7+ 3\alpha_8\\
   \beta_5&=2\alpha_1+4\alpha_2+ 5\alpha_3+  4\alpha_4+3\alpha_5+ 2\alpha_6+ \alpha_7+3\alpha_8\\
   \beta_6&=2\alpha_1+4\alpha_2+5\alpha_3+ 4\alpha_4+3\alpha_5+2\alpha_6+ \alpha_7+2\alpha_8\\
   \beta_7&=2\alpha_1+ 3\alpha_2+5\alpha_3+  4\alpha_4+ 3\alpha_5+ 2\alpha_6+\alpha_7+ 3\alpha_8\\
   \beta_8&=\alpha_1+ 3\alpha_2+5\alpha_3+  4\alpha_4+ 3\alpha_5+2\alpha_6+ \alpha_7+3\alpha_8\\
   \beta_9&=2\alpha_1+ 3\alpha_2+ 5\alpha_3+  4\alpha_4+3\alpha_5+2\alpha_6+\alpha_7+ 2\alpha_8\\
   \beta_{10}&=\alpha_1+ 3\alpha_2+5\alpha_3+  4\alpha_4+3\alpha_5+ 2\alpha_6+ \alpha_7+2\alpha_8\\
   \beta_{11}&=2\alpha_1+3\alpha_2+4\alpha_3+  4\alpha_4+ 3\alpha_5+ 2\alpha_6+ \alpha_7+2\alpha_8\\
   \beta_{12}&=\alpha_1+3\alpha_2+4\alpha_3+ 4\alpha_4+ 3\alpha_5+2\alpha_6+\alpha_7+2\alpha_8\\
   \beta_{13}&=2\alpha_1+ 3\alpha_2+4\alpha_3+  3\alpha_4+ 3\alpha_5+2\alpha_6+\alpha_7+ 2\alpha_8\\
   \beta_{14}&=\alpha_1+3\alpha_2+ 4\alpha_3+ 3\alpha_4+ 3\alpha_5+2\alpha_6+ \alpha_7+ 2\alpha_8\\
   \beta_{15}&=\alpha_1+ 2\alpha_2+ 4\alpha_3+  4\alpha_4+3\alpha_5+2\alpha_6+\alpha_7+ 2\alpha_8\\
   \beta_{16}&=2\alpha_1+3\alpha_2+ 4\alpha_3+  3\alpha_4+ 2\alpha_5+2\alpha_6+\alpha_7+ 2\alpha_8\\
   \beta_{17}&=\alpha_1+ 3\alpha_2+ 4\alpha_3+  3\alpha_4+ 2\alpha_5+ 2\alpha_6+\alpha_7+2\alpha_8\\
   \beta_{18}&=\alpha_1+ 2\alpha_2+ 4\alpha_3+  3\alpha_4+ 3\alpha_5+2\alpha_6+\alpha_7+ 2\alpha_8\\
   \beta_{19}&=2\alpha_1+3\alpha_2+ 4\alpha_3+  3\alpha_4+ 2\alpha_5+ \alpha_6+\alpha_7+ 2\alpha_8\\
   \beta_{20}&=\alpha_1+3\alpha_2+ 4\alpha_3+  3\alpha_4+ 2\alpha_5+ \alpha_6+ \alpha_7+2\alpha_8\\
   \beta_{21}&=\alpha_1+ 2\alpha_2+4\alpha_3+  3\alpha_4+ 2\alpha_5+ 2\alpha_6+ \alpha_7+2\alpha_8\\
   \beta_{22}&=\alpha_1+ 2\alpha_2+3\alpha_3+  3\alpha_4+3\alpha_5+ 2\alpha_6+\alpha_7+ 2\alpha_8\\
   \beta_{23}&=\alpha_1+ 2\alpha_2+ 4\alpha_3+  3\alpha_4+ 2\alpha_5+ \alpha_6+ \alpha_7+ 2\alpha_8\\
   \beta_{24}&=\alpha_1+ 2\alpha_2+ 3\alpha_3+  3\alpha_4+ 2\alpha_5+2\alpha_6+\alpha_7+ 2\alpha_8\\
   \beta_{25}&=\alpha_1+ 2\alpha_2+ 3\alpha_3+  3\alpha_4+ 2\alpha_5+ \alpha_6+ \alpha_7+2\alpha_8\\
   \beta_{26}&=\alpha_1+2\alpha_2+3\alpha_3+  2\alpha_4+ 2\alpha_5+2\alpha_6+\alpha_7+2\alpha_8\\
   \beta_{27}&=\alpha_1+ 2\alpha_2+ 3\alpha_3+  2\alpha_4+2\alpha_5+ \alpha_6+ \alpha_7+2\alpha_8\\
   \beta_{28}&=\alpha_1+2\alpha_2+3\alpha_3+  2\alpha_4+\alpha_5+ \alpha_6+\alpha_7+2\alpha_8
   \end{split}
  \]

\end{itemize}

\end{appendices}
\vfill\eject
\bibliographystyle{utphys}
\bibliography{ref}

\end{document}